# COZMO - A New Lightweight Stream Cipher

Rhea Bonnerji[0000-0002-5825-8800], Simanta Sarkar[0000-0002-4210-2764],
Krishnendu Rarhi[0000-0002-5794-215X], Abhishek Bhattacharya

School of Information Technology, Institute of Engineering & Management, Kolkata

**Abstract.** This paper deals with the merger of the two lightweight stream ciphers – A5/1 and Trivium. The idea is to make the key stream generation more secure and to remove the attacks of the individual algorithms. The bits generated by the Trivium cipher (output) will act as the input of the A5/1 cipher. The registers used in the A5/1 cipher will be filled by the output bits of the Trivium cipher. The three registers will then be connected to generate an output which will be our required key stream.

## 1 Introduction

Lightweight stream ciphers are used to reach high levels of security using only a small computing power. Stream encryption is the encryption of each letter one by one followed by the changing the encryption key after each letter. Lightweight stream ciphers have the advantage of having low cost hardware implementations as they have high throughput and low complexity. The idea is to come up with a lightweight stream cipher algorithm by modifying and merging existing cipher algorithms to make it secure to the attacks the stream ciphers were originally vulnerable to.Here, we are merging and making changes in Trivium and A5/1 algorithm to build a new lightweight stream cipher COZMO. In section 1.1 and 1.2 we are explaining how Trivium and A5/1 work. In section 2, we are explaining the working of COZMO. Then in section 3, we are discussing the statistical results in detail. Section 4 contains the conclusion of the proposal.

## Background:

### 1.1 Understanding the working of Trivium:

Trivium generates up to $2^{64}$ bits of key stream from an 80 bit secret key and an 80 bit initialization vector (IV). First the internal state of the cipher is initialised using the key and the IV. Then the state is updated repeatedly to generate the key stream bits. It consists of 3 interconnected non-linear feedback shift registers. The length of the registers are 93, 84 and 111 bits, respectively.Initialisation requires 1152 steps of the clocking before key stream is generated. There are XOR gates to XOR the ouputs we get from the various registers which is again fed back to the first one. The 91 and 92 bits are ANDed and used to give the feedback to the 94$^{th}$ bit. The 175 and 176 bits are ANDed and used to give the feedback to the 178$^{th}$ bit.

The 286 and 287 bits are ANDed and used to give feedback to the 288th bit. And the 288th bit is giving back the feedback to the 1st bit. [1],[5],

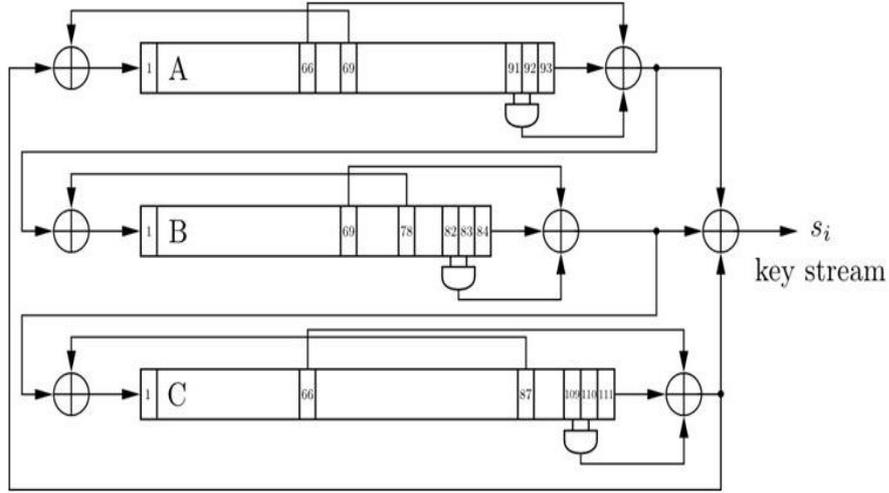

Fig. 1. Trivium

**Mathematical structure**

If we denote the internal state bits of the Trivium Model algorithm at time t as

$z(t)=(z_1(t), z_2(t), \ldots, z_{3n_3}(t))$, then the internal bits from time t to time t+ 1 can be expressed as follows

$$z(t + 1) = A \cdot z(t) + b(t) \quad (1)$$

Where $A = (a_{ij})_{3n_3 \times 3n_3}$ is the state-transition matrix of the algorithm with size $3n_3 \times 3n_3$

$$a_{ij} = \begin{cases} 1, & i=1, j= 3u_2, 3u_5, 3n_2 \\ 1, & i=3n_1+1, j= 3u_1, 3u_4 \\ 1, & i=3n_2+1, j= 3u_3, 3u_6 \\ 1, & 1<i\leq 3n_3, j =I - 1 \\ 0, & \text{otherwise} \end{cases} \quad (2)$$

$b(t)=(b_i(t))_{3n_3}$ is the nonlinear segment of the algorithm, which is treated as the vectors of bits

$$b_i(t) = \begin{cases} z_{3n_3-2}(t) \cdot z_{3n_3-1}(t), & i= 1 \\ z_{3n_1-2}(t) \cdot z_{3n_1-1}(t), & i= 3n_1 + 1 \\ z_{3n_2-2}(t) \cdot z_{3n_2-1}(t), & i= 3n_2 + 1 \\ 0, & \text{otherwise} \end{cases} \quad (3)$$

**1.2 Understanding the working of A5/1:**

A5/1 uses 3 shift registers of 19, 22 and 23 bits respectively. We have a 64 bit key and we'll load it in the three registers. Then we'll do some process to generate as many bits as we want to and we'll use those bits as the key stream and XOR it with the plain text to encrypt and the cipher text to decrypt. There is a majority vote function which will take 3 bits and find the majority of them. The three shift registers are loaded with the 64 bit key. There are three special positions in the registers – 8, 10 and 10 respectively. Take the majority of these three bits and check which of the registers are in majority. If register 1 is in majority then take the $13^{th}$, $16^{th}$, $17^{th}$ and $18^{th}$ bit and XOR them. If register 2 is in majority then XOR the bits in the $20^{th}$ and $21^{st}$ position. If register 3 is in majority then XOR the bits in the $7^{th}$, $20^{th}$, $21^{st}$ and $22^{nd}$ position. Only shift the XORed bits to the first position of the registers belonging in majority. The register which is not in majority will remain untouched. Then we have to XOR the bits in the last position of the registers and that is our required key stream bit. Repeating this process over and over again will give us our key stream. [2],[10]

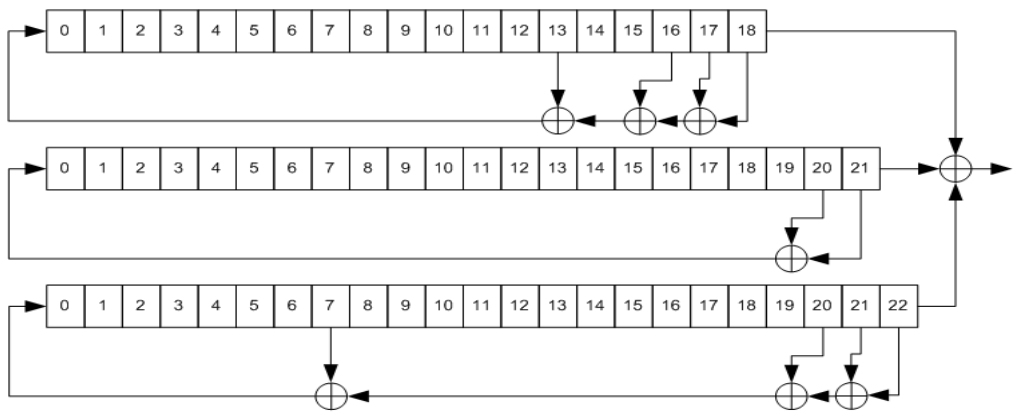

Fig. 2. A5/1

## 2 Proposed Methodology

The idea is to make the key stream even more secure by merging the two algorithms together. The Trivium cipher is initialised with a key and initialization vector and the A5/1 cipher is initialised with all 0's. First we will be generating a key stream using the Trivium algorithm. The generated key stream bits will be used as the input for the A5/1 cipher. There will be a clocking of 1216 (1152+64) cycles before the key stream starts getting generated. Let the three registers be called register A, B and C respectively. First we XOR the selected position bits in all the registers. For the first register A, the selected positions are the $13^{th}$, $16^{th}$ and $17^{th}$ and $18^{th}$ bits. For the second register B, the selected positions are the $20^{th}$ and $21^{st}$ bits. For the third register C, the selected positions are the $7^{th}$, $20^{th}$, $21^{st}$ and $22^{nd}$ bits. Next we check the majority function and find the two registers that are in majority. Next,we shift the bit in the last position to the first position by using right shift by one place only for the registers in majority. Once that is done, we will use the XOR bits. If register A is in majority and B isn't then discard the XORed bit from register A. Since register C is in majority, the XORed bit of register B will replace the bit in the first position of register C. Register A which is also in majority will have its first bit replaced by the XORed bit of the register C which is XORed with the bit generated by Trivium. The final key stream is the XOR of the three bits in the last positions of the three registers.

## 2.1 Algorithm:

Step 1:

Initialise all the bits of the Trivium cipher with a key and an initialisation vector.

Initialise all the bits of A5/1 cipher with 0.

Step 2:

Clock 1216 (1152+64) cycles before generating the first key stream. Then generate key stream using Trivium. This key stream will be used as the input bits for A5/1.

Step 3:

XOR the $13^{th}$, $16^{th}$, $17^{th}$ and $18^{th}$ bits in register A. XOR the $20^{th}$ and $21^{st}$ bits in register B. XOR the $7^{th}$, $20^{th}$, 21t and $22^{nd}$ bits in register C.

Step 4:

Check the majority function and find the resisters that are in majority

Step 5:

For the registers in majority, right shift the bits by one place and replace the first bit of thregister with the XORed bit (r0 byp3, r19 by p1, r41 by p2).

Step 6:

Discard the XORed bit for the registers not in majority.

Step 7:

XOR the bits in the last position the registers.

Step 8:

Repeat steps 4-7.

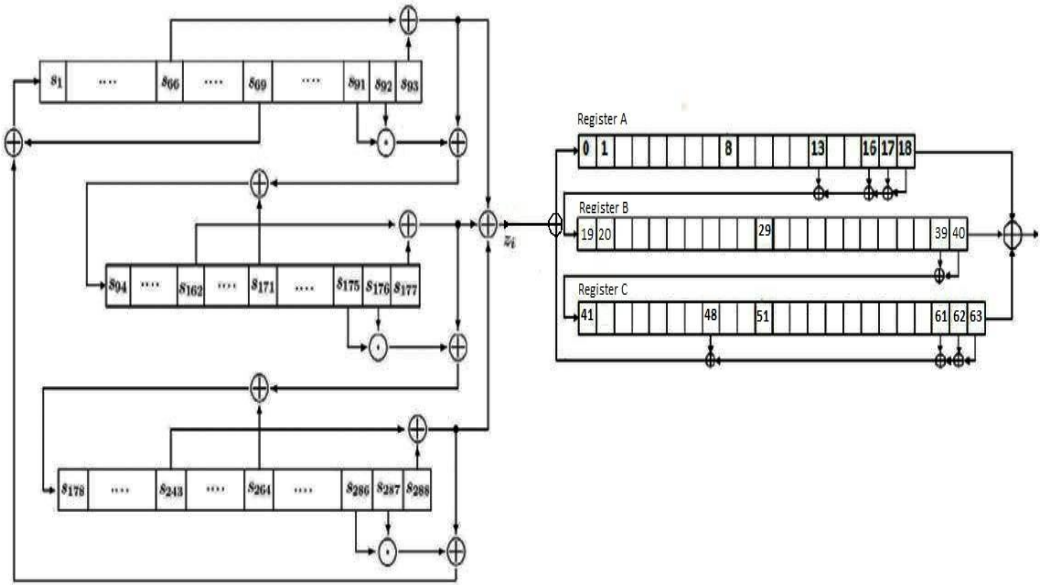

Fig. 3. COZMO

**2.3 Pseudo Code:**

**Key = (**K1, . . . , K80**)**

**IV =** (IV1, . . . , IV80)

(s1, s2, . . . , s93) ← (K1, . . . , K80, 0, . . . , 0)

(s94, s95, . . . , s177) ← (IV1, . . . , IV80, 0, . . . , 0)

(s178, s279, . . . , s288) ← (0, . . . , 0, 1, 1, 1)

Register A:

    (r0, r2, . . . , r18) ← (0,0, . . . , 0)

Register B:

    (r19, r20, . . . , r40) ← (0,0, . . . , 0)

Register C:

    (r41, r43, . . . , r63) ← (0,0, . . . , 0)

**Operations:**

⊕    : bit-wise exclusive OR

&    : bit-wise AND

**Variables:**

$z_i$ : the $i^{th}$ bit generated by trivium.

L : $r_8$ bit of register A

M : $r_{29}$ bit of register B

N : $r_{51}$ bit of register C

$t_i$ : The keystream bit generated at the $i^{th}$ step.

**Function:**

maj(L, M, N) = (L&M) $\oplus$ (M&N) $\oplus$ (L&N)

A complete description is given by the following pseudo-code:

For i = 1 to N do

    $t_i \leftarrow r_{18} \oplus r_{40} \oplus r_{63}$

    $p_1 \leftarrow r_{13} \oplus r_{16} \oplus r_{17} \oplus r_{18}$

    $p_2 \leftarrow r_{39} \oplus r_{40}$

    $p_3 \leftarrow r_{48} \oplus r_{61} \oplus r_{62} \oplus r_{63} \oplus z_i$

    If $s_8$ equal to maj(L, M, N) Then

        (r0, r1, r2, . . . , r18) $\leftarrow$ (p3, r0, . . . , r17)

    ElseIf $s_{29}$ equal to maj(L, M, N) Then

        (r19, r20, . . . , r40) $\leftarrow$ (p1, r19, . . . , r39)

    ElseIf $s_{52}$ equal to maj(L, M, N) Then

        (r41, r42, . . . , r63) $\leftarrow$ (p2, r41, . . . , r62)

    EndIf

End for

## 3 Results and Discussion

To test the randomness property of the key bit generator following statistical tests were conducted using the NIST Statistical Test Suite.[17]

**Frequency Test:** The test was conducted to check if the sequence is truly random. Randomness of the sequence is determined by the number of ones and zeros in the sequence, which is expected to be approximately equal for a truly random sequence.

**Cumulative Sums Test:** The focus of this test is the maximal excursion from zero of the cumulative sum of the partial sequences defined by the cumulative sum of adjusted (-1, +1) digits in the sequence. For a random sequence, the cumulative sum of the partial sequences should be near zero.

**Approximate Entropy Test:** The focus of this test is the frequency of each and every overlapping m-bit pattern. The purpose of the test is to compare the frequency of overlapping blocks of twoconsecutive/adjacent lengths (m and m+1) against the expected result for a random sequence.

**Linear Complexity:** This test is performed to determine whether or not the sequence is complex enough to be considered random. Random sequences are characterized by a longer feedback register.

**Serial Test:** The purpose of this test is to determine whether the number of occurrences of the 2m m-bit overlapping patterns is approximately the same as would be expected for a random sequence.

**Longest Runs of Ones:** The purpose of this test is to determine whether the length of the longest run of ones within the tested sequence is consistent with the length of the longest run of ones that would be expected in a random sequence.

**Runs Test:** The runs test is conducted to determine whether the number of runs of ones and zeros of various lengths is as expected for a random sequence. In particular, this test determines whether the oscillation between such substrings is too fast or too slow.

From the results of the tests tabulated below we can conclude that at 1% level of significance null hypothesis is accepted since the p-values are greater than 0.01. In other words, the data has passed all the above-mentioned tests proving randomness property of the key bit generator.

| Statistical Test | p- value | Success/failure |
|---|---|---|
| Frequency | 0.534146 | Success |
| Cumulative Sums | P1-0.122325 P2-0.350485 | Success |
| Approximate Entropy | 0.004301 | Success |
| Linear Complexity | 0.213309 | Success |
| Serial | P1-0.004301 P2- 0.017912 | Success |
| Longest Run of Ones | 0.350485 | Success |
| Runs | 0.911413 | Success |

Table 1: The Results of Statistical Analysis for A5/1

| Statistical Test | p- value | Success/failure |
|---|---|---|
| Frequency | 0.534146 | Success |
| Cumulative Sums | P1- 0.213309 P2- 0.350485 | Success |
| Approximate Entropy | 0.534146 | Success |
| Linear Complexity | 0.035174 | Success |
| Serial | P1- 0.911413 P2- 0.534146 | Success |
| **Longest Run of Ones** | 0.739918 | Success |
| **Runs** | 0.350485 | Success |

Table 2: The Results of Statistical Analysis for Trivium

| Statistical Test | p- value | Success/failure |
|---|---|---|
| **Frequency** | 0.911413 | Success |
| **Cumulative Sums** | P1-0.991468<br>P2-0.739918 | Success |
| **Approximate Entropy** | 0.350485 | Success |
| **Linear Complexity** | 0.066882 | Success |
| **Serial** | P1-0.213309<br>P2-0.534146 | Success |
| **Longest Run of Ones** | 0.534146 | Success |
| **Runs** | 0.122325 | Success |

Table 3: The Results of Statistical Analysis for COZMO

## 4 Conclusion

We want to propose a new lightweight stream cipher[7] because of their higher speed, efficiency and its easy implementations on applications which require plain texts of unknown length. In this paper we merged and made some changes to the algorithms of two already existing lightweight stream ciphers – Trivium and A5/1 and proposed a new lightweight stream cipher with good key randomness as key randomness is a major issue is stream ciphers. The generated key stream is more secure compared to the key streams generated by the individual ciphers. Our algorithm produces sequences which are stronger in statistical properties than the A5/1 and Trivium. The implementation of this algorithm is also very feasible and easy to put to use.